\documentstyle[12pt,epsf,epsfig]{article}
\def\ga{\mathrel{\raise.3ex\hbox{$>$\kern-.75em\lower1ex\hbox{$\sim$}}}}
\def\la{\mathrel{\raise.3ex\hbox{$<$\kern-.75em\lower1ex\hbox{$\sim$}}}}

\renewcommand{\thefootnote}{\fnsymbol{footnote}}

\newcommand{\nc}{\newcommand}

\nc{\non}{\nonumber}
\nc{\hc}{\hbox {h.c.}} \nc{\re}{\hbox {Re}} 
\nc{\mev}{\hbox {MeV}} \nc{\gev}{\;\hbox {GeV}} \nc{\tev}{\;\hbox {TeV}}
\def\lsim{\mathrel{\raise.3ex\hbox{$<$\kern-.75em\lower1ex\hbox{$\sim$}}}}
\def\gsim{\mathrel{\raise.3ex\hbox{$>$\kern-.75em\lower1ex\hbox{$\sim$}}}}

\nc{\prd}[3]{{\it Phys.\ Rev.}\ {{\bf D{#1}} (#2), #3}}
\nc{\prl}[3]{{\it Phys.\ Rev.\ Lett.}\ {{\bf {#1}} (#2), #3}}
\nc{\plb}[3]{{\it Phys.\ Lett.}\ {{\bf B{#1}} (#2), #3}}
\nc{\npb}[3]{{\it Nucl.\ Phys.}\ {{\bf B{#1}} (#2), #3}}
\nc{\ptp}[3]{{\it Prog.\ Theor.\ Phys.}\ {{\bf {#1}} (#2), #3}}
\nc{\zfp}[3]{{\it Z.\ Phys.}\ {{\bf C{#1}} (#2), #3}}
\nc{\epj}[3]{{\it Eur.\ Phys.\ J.}\ {{\bf C{#1}} (#2), #3}}
\nc{\mpla}[3]{{\it Mod.\ Phys.\ Lett.}\ {{\bf A{#1}} (#2), #3}}
\nc{\rmp}[3]{{\it Rev.\ Mod.\ Phys.}\ {{\bf {#1}} (#2), #3}}
\nc{\ijmpa}[3]{{\it Int.\ J.\ of\ Mod.\ Phys.}\
               {{\bf A{#1}} (#2), #3}}
\nc{\Lsp}{\;\;\;\;\;\;\;\;\;\;}  \nc{\LLLsp}{\lspace \lspace}
\nc{\lsp}{\;\;\;\;\;\;}
\nc{\spac}{\;\;\;}
\nc{\noi}{\noindent}
\nc{\beq}{\begin{equation}}   \nc{\eeq}{\end{equation}}
\nc{\bea}{\begin{eqnarray}}   \nc{\eea}{\end{eqnarray}}
\nc{\baa}{\begin{array}}      \nc{\eaa}{\end{array}}
\nc{\bit}{\begin{itemize}}    \nc{\eit}{\end{itemize}}
\nc{\ben}{\begin{enumerate}}  \nc{\een}{\end{enumerate}}
\nc{\bce}{\begin{center}}     \nc{\ece}{\end{center}}
\begin{document}

\baselineskip=16pt
\rightline{UCD-0217}
\rightline{December 2002}  
\begin{center}

\vspace{0.5cm}

\large {\bf The radion and the perturbative metric in RS1}
\vspace*{5mm}
\normalsize

{\bf  Manuel Toharia}\footnote{
\tt mtoharia@physics.ucdavis.edu}

\smallskip 
\medskip 
 
{\it Physics Department, University of California,\\  
Davis, CA 95616-8677, USA} 

\smallskip 
\end{center} 
\vskip0.6in 
 
\centerline{\large\bf Abstract}

\vspace{.5cm}

{\footnotesize

We calculate the linearized metric perturbations in the five dimensional two-brane model of Randall and Sundrum. In a carefully chosen gauge, we write down and decouple Einstein equations for the perturbations and get the final and simple perturbative metric ansatz. This ansatz turns out to be equal to the linear expansion of the metric solution of Charmousis et al. \cite{rubakov}. We show that this ansatz, the metric ansatz of Boos et al. \cite{boos} and the one of Das and Mitov \cite{das} are not incompatible, as it appears on the surface, but completely equivalent by an allowed gauge transformation that we give.\\

\vspace{.5cm}



}


\renewcommand{\thefootnote}{\arabic{footnote}}
\setcounter{footnote}{0}

\section{Introduction}

We consider the Randall-Sundrum background scenario \cite{rs}, where we have an extra dimension that is compact and verifies a $Z_2$ symmetry. All matter fields are supposed to live on the two branes of the scenario. In order for the static background of this setup to solve Einstein equations, the tensions on the branes must be fine-tuned (they have to be related with one another, and with the non-zero, negative Cosmological Constant of the Bulk).
This scenario was introduced originally to explain the hierarchy problem between Planck and Electroweak scales. The warp factor appearing in the 5-dimensional metric of the model allows to generate this huge hierarchy in a `natural way'.

But when we add dynamical degrees of freedom to the static background of the setup, the existence of the two branes will have important consequences.

We are interested in weak perturbations of the metric, and these will in general include a tensor field, a vector field (the $({}_{\mu5})$ part of the metric) and a scalar field ($({}_{55})$ part of the metric).

In the weak field approach, we always have a gauge freedom on the metric perturbations, and we can ask ourselves what can we gauge away by fixing appropriately the gauge. With two branes, it turns out that the vector perturbations (${}_{\mu5}$) can be completely gauged away, and the 5-d scalar perturbation can be turned into a 4-d scalar perturbation.

This was done in \cite{rubakov} and \cite{garriga} by using the so-called `bent-brane' procedure (see also \cite{pilo}). This procedure in general terms, uses two coordinate frames, in which one of the branes is not at fixed-y position. In these frames, we can always gauge away vector and scalar perturbations, but when switching to a frame where both branes are at fixed-y, a massless 4-d scalar field appears and is identified as the radion.

There has been some discrepancy in the literature about this method \cite{arefeva,boos} and we therefore propose instead, to carry out a gauge fixing procedure that will maintain always the two branes fixed, and show that the linearized version of \cite{rubakov} is actually correct and consistent. The metric ansatz calculated in \cite{boos} and originally introduced to correct the problems of the `bent brane' method, will turn out to be completely equivalent with our result, the difference residing on a simple gauge transformation, that we will write down.

We will not address here the issue of the stabilization of the radion and its consequences on its couplings. This issue comes from the fact that the inter-brane distance of the static background is not set by any scale. This `scale-invariance' of the background, is ultimately responsible for the masslessness of the radion. The problem, among others, is that for the model to solve the hierarchy problem, we need to fix the interbrane distance to some specific value. This fixing, or stabilization, would induce a potential for the radion, that would become massive. Some specific mechanisms have been proposed and studied in the literature \cite{gw,tanaka,csaki,maxim} and we do not intend to study here any of them in the context of our gauge fixing process.

But before anything, let's first review the static background setup of Randall and Sundrum.

Through the rest of this paper, capital Latin letters $(A,B,C,..)$ run from 1 to 5 and Greek letters $(\mu,\nu,..)$ run from 1 to 4, and primes ($\ {'}\  $) denote derivatives with respect to the extra dimension $y$.

\section{Randall-Sundrum setup}

The background action of the model is
\bea
{\cal S}= - M_*^3 \int d^5x\ \sqrt{g_5}{\cal R} - \int d^5x \sqrt{g_5}\ \Lambda  -  \int d^5x  \sqrt{\hat{g}_1 }\ \lambda_1\ \delta({y}) - 
\int d^5x \sqrt{\hat{g}_2}\ \lambda_2\ \delta({y-r_0})\ \ \ 
\eea 
and we want to find a solution for the static background metric $\bar{g}_{AB}$ of the form
\bea 
ds^2 &=& e^{-2\sigma(y)}\ \eta_{\mu \nu } dx^\mu dx^\nu + dy^2 \label{bckgrnd}
\eea
where $\eta_{\mu\nu}$ is chosen in the convention $(-+++)$.

In general, the induced metrics $\hat{g}_1$ and $ \hat{g}_2$ on each brane have the form  
\bea
\hat{g}_{\mu\nu}(x) = \bar{g}_{AB}(X(x))\ \partial_\mu X^A\ \partial_\mu X^B
\eea
where $X^A(x)$ describes the position of the brane in a 5-dimensional reference frame with respect to the position $x$ on the brane.

We will work in a frame in which $\ \ \partial_\mu X^A\ \partial_\mu X^B = \delta_\mu^A \delta_\nu^B$ so that the induced metrics are simply of the form $ \hat{g}_{\mu \nu}(x) = g_{\mu \nu}(x) $. In this frame, the two branes sit on a fixed y-position.

The Einstein equations for the background configuration give then
\bea
{2 \sigma'^2 -  \sigma''} &=& { 1\over 6M_{*}^3} \left[-\Lambda - {\lambda_1 }\ \delta(y) - {\lambda_2}\ \delta(y-r_0) \right]\label{condeltas}\\
&{\rm and} &\nonumber\\
{ \sigma'^2 } &=& -{ \Lambda \over 12 M_{*}^3}\label{sindeltas} 
\eea
Now eq. (\ref{sindeltas}) tells us that $\sigma'^2$ is constant (and, therefore, continuous). But $\sigma'$ is not necessarily continuous, since 
\bea
\sigma' = \pm k 
\eea
If we choose $\sigma'(\epsilon) = k >0 \ $, which in turns requires that $\sigma'(-\epsilon) = -k$ (with $\epsilon$ being an arbitrarily small number), we get
\bea
2k ={ 1\over 6M_{*}^3} \lambda_1 \hspace{1cm} {\rm and} \hspace{1cm} 
2k =-{ 1\over 6M_{*}^3}\lambda_2
\eea

So that finally, we have the well-known relations between brane tensions and bulk cosmological constant, that is found to be negative.
\bea 
\lambda_1 = -\lambda_2\ =\ {12 k M_{*}^3} \hspace{3cm} \Lambda = {12 k^2 M_{*}^3}
\eea

These relations actually represent a forced fine-tuning of the model since the three terms are in principle of different nature and origin.

The warp function is\footnote{To be more precise and to cover the complete circle (and therefore include both branes), we should write \cite{gundanbog}, $\ \sigma(y)= k \left[ y (2 \theta(y) - 1 ) - 2 (y-r_0) \theta(y-r_0)\right] + c\ $. This makes explicit the presence of the second brane at $y=r_0$, and ensures that $\sigma'$ makes a jump on the two branes.}
\bea
\sigma(y) = k |y| + c \ \ \ \ { \rm for } \ \ -r_0 < y < r_0
\eea 
where c is some constant of integration that we will chose in the manner that follows.

If we live in the negative tension brane at $y=r_0\ $, we will set the constant c to $- k r_0$ so that the induced metric in our brane has the good old Minkowski form, and no field redefinitions will be necessary \cite{rubrev}. If we live in the brane at y=0, then we will set $c=0$ for the same reason.

We assume our brane is the one located at $y=r_0$ and therefore write
\bea
ds^2= e^{-2 (k |y| -  k r_0 )} \eta_{\mu \nu} + dy^2 
\eea

\section{Perturbations around the Background}

In what follows, we will not use the explicit form of the warp function $\sigma(y)$. Instead, we will carry out everything in terms of a generic $\sigma(y)\ $ function, having in mind its fundamental properties (see eq. (\ref{condeltas}) and (\ref{sindeltas}))
\bea
\sigma''(y) &=& {1\over 6 M_{*}^3} \left[ \lambda_1\ \delta(y) + \lambda_2\ \delta(y-r_0)\right]\\
{\sigma'}^2(y)&=&\ \  k^2\\
\sigma'(-\epsilon)\ =\ - \sigma'(\epsilon)\ & {\rm and}  &\  \sigma'(r_0 -\epsilon)\ =\ - \sigma'(r_0+\epsilon)
\eea

This will ensure that we are always keeping track of the $S_1/Z_2$ properties of the fields in the setup, and that every time we encounter a $\sigma''\ $, we know we are handling the information coming from the branes. It also allows us to work in a general manner without having to solve first in the bulk and then impose boundary field equations.

\subsection{Gauge freedom equations}

We define the metric perturbations $h_{AB}$ as
\bea
g_{AB} = \bar{g}_{AB} + \epsilon\ h_{AB}
\eea
In the weak field approximation, when we take the inverse of the metric, we will only keep terms up to some order in $\epsilon$.

The functions $h_{AB}(x,y)$ satisfy the symmetry conditions:
\bea
h_{\mu\nu}(x,-y)&=&h_{\mu\nu}(x,y)\\
h_{\mu 5}(x,-y)&=&- h_{\mu 5}(x,y)\\
h_{5 5}(x,-y)&=& h_{5 5}(x,y)
\eea
There is a gauge freedom on the metric perturbations that maintains the first order lagrangian invariant, and it is related to infinitesimal coordinate transformations.
\bea
x^{new}_A &=& x_A + \xi_A(x,y)\nonumber\\
\Rightarrow\ \ \ \ h^{new}_{AB}&=& h_{AB} - \nabla_A \xi_B(x,y) - \nabla_B \xi_A(x,y) \label{ggfreedom}
\eea 
where $\nabla_A$ is the covariant derivative.

We can calculate explicitly these gauge equations for our simple background metric.
\bea
h^{new}_{\mu\nu}(x,y) &=& h_{\mu\nu}(x,y) - \partial_\mu \xi_\nu - \partial_\nu \xi_\mu + 2 \sigma'\ \xi_5\ \bar{g}_{\mu\nu} \label{gmunu}\\ 
h^{new}_{\mu 5}(x,y) &=& h_{\mu 5}(x,y) - \partial_\mu \xi_5 - e^{-2\sigma} \left( e^{2\sigma} \xi_\mu \right)'\label{gmu5}  \\
h^{new}_{55}(x,y) &=& h_{55}(x,y) - 2\ \partial_5 \xi_5 \label{g55}
\eea
It is crucial to realize that the functions $\xi_A$ satisfy the symmetry conditions
\bea
\xi_\mu(x,-y) &=& \xi_\mu(x,y) \\
\xi_5(x,-y) &=& - \xi_5(x,y)
\eea
and the boundary conditions \cite{boos,arefeva,kakushadze}
\bea
\xi_5(x,0)\hspace{.5cm}=\hspace{.3cm} \xi_5(x,r_0)\hspace{.6cm} &=& 0\\
 \left(e^{2\sigma} \xi_\mu\right)'\Big|_{y=0} = \left(e^{2\sigma} \xi_\mu\right)'\Big|_{y=r_0}   &=& 0
\eea
Now, we want to write down the equations of motion of the metric perturbations from Einstein equations. But the task will be simplified if we choose a gauge in which the number of perturbations is reduced. 


\subsection{Choosing the gauge}

We will first try to fix the gauge for the field $h_{55}(x,y)$ but, we insist, making sure we maintain the branes at fixed-y. This means that we need to ensure that the function $\xi_5(x,y)$ that we will choose vanishes on each brane location.

We propose
\bea
\xi_5(x,y)= {1\over4} \int^y_{-y}dY h_{55}(x,Y)\ -\ {y\over 4r_0} \int^{r_0}_{-r_0} dY h_{55}(x,Y)\ - {1\over2}\ \sigma'\ J[\sigma(y)]\ r(x)
\label{xi5}
\eea
where $r(x)$ is any function of the $4-d$ coordinates $x$ and where $J[\sigma(y)]$ is some unknown function with the boundary conditions $J(0)=J(r_0)=0$, and that will be determined later. This choice ensures that $ \xi_5(x,y)$ vanishes on the two brane locations at $y=0$ and $y=r_0$. Moreover, we see that $ \xi_5(x,y)=-\xi_5(x,-y) $, in agreement with the orbifold symmetry.

We plug this in (\ref{g55}) to get
\bea 
h^{new}_{55}(x,y) &=& {1\over 2r_0} \int^{r_0}_{-r_0} dY h_{55}(x,Y)\ + \sigma'^2 {\partial J[\sigma]\over \partial \sigma}\  r(x) + \sigma'' J[\sigma(y)] r(x) \ 
\eea
The $\sigma''$ term is identically zero. And we can still fix our choice more by asking that
\bea \alpha_0\ r(x) = {1\over 2 r_0} \int^{r_0}_{-r_0} dY h_{55}(x,Y)\label{ffix}\eea
with $\alpha_0 \neq 0$, being some constant that we can fix later.

Our gauge-fixed $h_{55}$ metric perturbation becomes
\bea 
h^{new}_{55}(x,y) = F(y)\ r(x)
\eea
with the function $F(y)$ defined as
\bea
\hspace{2cm}F(y) = \left( \alpha_0 +  \sigma'^2 {\partial J \over \partial \sigma}\ \right)\hspace{1cm} {\rm with} \ \ \ \ \ J(0)=J(r_0)=0\label{Fdef}
\eea

We see that it was not possible to gauge away totally the perturbation $h_{55}$ (because the boundary conditions impose $F(y) \neq 0$). 

In the case of a one brane scenario, the last two terms in eq. (\ref{xi5}) would not be needed, and gauging away the perturbation $h_{55}$ would have been possible (the translational `vibration' of one brane is unphysical). But, when we add a second brane, we impose an extra boundary condition that forces us to keep some dynamical function. It was identified as the radion \cite{arkani,sundrum} and it keeps track of the relative `vibrational motion' of the two branes \cite{rubakov}.   

The next step is to gauge away the $h_{\mu5}$ perturbation which turns out to be possible \cite{boos,kakushadze,arefeva}, and we even keep some residual gauge freedom on the tensor perturbations.
\bea
h^{new}_{\mu\nu}(x,y) &=& h_{\mu\nu}(x,y) - e^{-2\sigma(y)}\ \partial_\mu V_\nu(x) - e^{-2\sigma(y)}\ \partial_\nu V_\mu(x) \label{gaugefreedom} 
\eea

Our gauge choice can be summarized by  
\bea
h_{55}(x,y) &=& F(y)\ r(x)\\
h_{\mu 5}(x,y) &=& 0
\eea 
and the corresponding metric ansatz is
\bea
ds^2 = \Big( \bar{g}_{\mu \nu}\ + \epsilon\ e^{-2\sigma} j_{\mu \nu}(x,y)  \Big)\  dx^\mu dx^\nu + (1 +  \epsilon\ \ F(y)\ r(x)\ )\  dy^2 \label{metric}
\eea
where $\bar{g}_{\mu \nu} \ = e^{-2\sigma} \eta_{\mu\nu}$ is the background metric that we will use to raise and lower indices.

We have used the notation $j_{\mu \nu}(x,y) = e^{2\sigma} h_{\mu \nu}(x,y)$ instead of simply $h_{\mu\nu}$ because the equations of motion turn out to be simpler for the $j_{\mu\nu}$ field, but also because the residual gauge freedom on the tensor perturbation seems to point to this parametrization (see eq. (\ref{gaugefreedom})). 

The problem now is that the equations of motion for these perturbations will turn out to be coupled and we need to decouple them to get the correct physical fields.

\subsection{Decoupling the equations of motion}
We intend to write and decouple the field equations for the perturbations in the absence of any matter field.
 
The perturbed Einstein equations give a visually complicated set of equations that we will divide in 4 sets: the ${}{(\mu \nu)}$ equations, the ${}{(\mu5)}$ equations, the ${}{(55)}$ equation and the contracted equation $[({}^\mu_{\ \mu}) - (55)]$ or (${\footnotesize {}^A_{\ A}}$), which will help us in the decoupling process.

\noindent For the ${}{(\mu \nu)}$ equations we get
\bea
&&\hspace{-2cm} {1\over2} e^{4 \sigma} \left( j'_{\mu\nu} e^{4 \sigma} \right)'+ {F e^{2\sigma} \over 2}\partial_\mu \partial_\nu r  +\ {1 \over2} \ (\partial^\alpha \partial_\mu j_{\alpha\nu} + \partial^\alpha \partial_\nu j_{\alpha\mu} - \partial^\alpha \partial_\alpha j_{\mu\nu} - \partial_\mu \partial_\nu j^\alpha_{\ \alpha})      \nonumber\\
&&\hspace{.1cm} +\ \bar{g}_{\mu\nu}\ \Big(- {F e^{2\sigma} \over 2} \bar{g}^{\alpha \beta}\partial_\alpha \partial_\beta r + {3\over 2} e^{6\sigma} r  \left(  F \sigma' e^{-4\sigma} \right)' +\ 2 \sigma' j'_{\alpha \beta} \bar{g}^{\alpha \beta}       \nonumber \\
&&\hspace{2.5cm} - {1 \over 2} \bar{g}^{\alpha \beta} j''_{\alpha \beta}-\ {1 \over 2} \{ \partial^\alpha \partial^\beta j_{\alpha \beta} - \partial^\alpha \partial_\alpha j^\beta_\beta \}\ \ \ \ \  \Big) \  = \  0 \label{munu}
\eea
for the ${}{(\mu5)}$ equation:
\bea
e^{-2\sigma} \bar{g}^{\alpha\beta} (-\partial_\mu j_{\alpha\beta} + \partial_\alpha j_{\mu\beta})' + 3 F  \sigma' \partial_\mu r \ = 0 \label{mu5}
\eea
for the ${}{(55)}$ equation
\bea
- \left( \partial^\mu \partial^\nu j_{\mu \nu}\ -\ \partial^\mu \partial_\mu j_{\ \nu}^\nu \right) +\ {3} \sigma'\  \bar{g}^{\mu\nu}\ j'_{\mu\nu}  + 12 F e^{2\sigma}\sigma'^2 r  =\ 0 \label{55}
\eea
and the contracted equation (${\footnotesize {}^A_{\ A}}$)
\bea
\hspace{-1cm} &&-{1 \over 2 } e^{2\sigma}\bar{g}^{\mu\nu}\left( e^{-2\sigma}j'_{\mu \nu} \right)'-\ {1\over 2} e^{2\sigma}F \bar{g}^{\mu\nu} \partial_\mu \partial_\nu r -\ \left(F \sigma' e^{-2 \sigma}\right)' \ 2 e^{4\sigma}r\ =\ 0\hspace{.3cm}
\eea 
We immediately see that these equations are coupled differential equations of motion, and to extract the physical fields we need to separate them in independent equations for each field. To decouple them, we will try the redefinition
\bea
j_{\mu\nu} = H_{\mu\nu} - G({y})\ \eta_{\mu\nu}\ r \label{redef}
\eea

By carefully analyzing the new equations that we get, we find that the double requirement 
\bea G^{'}= F \sigma' \hspace{1cm} {\rm and} \hspace{1cm} {F\over 2}=  G \label{defG}\eea
on our yet unknown functions $G(y)$ and $F(y)$ is the only one that decouples the equations.

We can then solve for $G(y)$ and $F(y)$:
\bea
 G(y) = \alpha\ e^{2\sigma(y)}\ = {F(y) \over 2}
\eea
with $\alpha$ an integration constant that we will arbitrarily set to 1.

To completely decouple the equations we also need to partially fix the residual gauge freedom that we still had for the field $H_{\mu\nu}$ (see eq. (\ref{gaugefreedom})) to finally get the decoupled equations of motion
\bea
\bar{g}^{\mu\nu} \partial_{\mu} \partial_{\nu} r = 0\label{boxr}
\eea
which is the equation of motion of a massless 4-dimensional scalar field.

and
\bea
 {1\over 2}e^{4\sigma} \left(e^{-4\sigma} H'^{\ TT}_{\mu\nu}\right)' - \ {1 \over2} \bar{g}^{\alpha\beta} (  \partial_\alpha \partial_\beta H^{\ TT}_{\mu\nu} ) \ =\ \  0
\eea 
which is the equation of motion of a 2-tensor field in the T-T gauge, i.e., such that
\bea
 H^{\mu}_{\ \mu} &=&  0\hspace{2cm}  {\rm and} \hspace{2cm}\partial^\alpha H_{\alpha\mu} = 0 
\eea
and we still have some residual gauge freedom on $H^{\ TT}_{\mu\nu}$
\bea
H^{TT^{new}}_{\mu\nu} &=& H^{TT}_{\mu\nu} -  \partial_\mu \chi_\nu(x) - \partial_\nu \chi_\mu(x) \hspace{1.3cm} {\rm with}\ \   \Box \chi_\mu = \partial^\mu \chi_\mu=0   
\eea

So, we have decoupled the equations of motion by redefining the tensor perturbation and this has fixed the function $F(y)$ that was part of our gauge fixed metric (eq. (\ref{metric})). It also fixed the function $G(y)$ that appeared in the tensor field redefinition (eq. (\ref{redef})).

The radion turns out to be a massless 4-dimensional scalar field, and we have proved that we can choose the 4-dimensional T-T gauge on the tensor perturbations.\\

Before we end this section, we remind the reader that the function $F(y)$ should verify some boundary conditions that we need to check for our solution to be consistent. 

We have
\bea
F(y) = \left( \alpha_0 +  \sigma'^2 {\partial J \over \partial \sigma}\ \right)\hspace{1cm} {\rm with} \ \ \ J(0)=J(r_0)=0
\eea
We need to integrate this equation and make sure that with $\ F(y)=2 e^{2\sigma(y)}$, the function $J(y)$ vanishes at $y=0$ and $y=r_0$. 

We get
\bea 
J(y) = {1\over k^2} e^{2\sigma(y)} - {\alpha_0\ \sigma(y)\over k^2} + \alpha_1\label{J}
\eea
where $\alpha_1$ is a constant of integration, and $\alpha_0$ is some constant that we can fix at our convenience. 
Obviously, a solution for the boundary problem exists: 
\bea
\alpha_1= -{1\over k^2} \hspace{1cm} {\rm and} \ \ \ \alpha_0= {1\over k r_0} (1 - e^{-2 k r_0} )\label{constants}
\eea
and, therefore, our solution for $F(y)$ is consistent, proving that the process of decoupling was successful.\\

Let us then write down our simple ansatz, which decouples the equations of motion
\bea
\hspace{-1cm}ds^2 = \left( \bar{g}_{\mu \nu} + \epsilon\ e^{-2\sigma(y)} H^{\ TT}_{\mu\nu}(x,y) - \epsilon\  \eta_{\mu\nu}\  r(x)  \right) dx^\mu dx^\nu + \left(1 +  2\ \epsilon\ e^{2\sigma(y)}\ r(x) \right)  dy^2 
\eea

This ansatz turns out to be equal to the linear expansion of the ansatz in \cite{rubakov} that was derived using a different approach (see also \cite{pilo}). 

Finally, we must mention that this radion that appears in the metric ansatz is not canonically normalized in the 4-dimensional sense, its kinetic term in the effective 4-d lagrangian not being canonical. A canonical normalization of the radion in the framework of the linear expansion of the metric of \cite{rubakov} (and, therefore, ours) can be found in \cite{csaki}.  

The process of extracting the effective 4-d lagrangian for the tensor perturbations can be found in \cite{boos}.

\subsection{Gauge transformed ansatz}

We have decoupled Einstein equations by fixing the gauge and using a simple redefinition of the tensor perturbations. So, starting from our ansatz, we are now going to define the gauge transformation
\bea
\xi_\mu(x,y) &=& -\ {1\over2}\ e^{-2\sigma(y)} \left( \int_0^y dY\ \sigma' e^{2\sigma(Y)}  J(Y)\  \right)  \ \    \partial_\mu r(x)\  \non \\
{\rm and}\hspace{3cm} &&\label{gauge}\\
\xi_5(x,y) &=&\ {1\over 2}\ \sigma' J(y) \ r(x) \non
\eea 
with $J(y)$ given in (\ref{J}) and (\ref{constants}).

From the gauge invariance equations (\ref{ggfreedom}), we can get the physical metric perturbations in this new gauge 
\bea
\hspace{-1cm}h^{new}_{\mu\nu}(x,y) &=& h_{\mu\nu}(x,y)\ -\  e^{-2\sigma(y)} \eta_{\mu\nu}\ \alpha_0\ \left({1\over \alpha_0} +   \sigma(y)\right)\  r(x) \non\\
&&\hspace{1.2cm}+\ \left[  {\alpha_0 \over 2 k^2} \left({1\over2} - {1\over \alpha_0} - \sigma(y) + {1\over 2 \alpha_0} e^{2\sigma(y)}\right) + C_0 e^{-2\sigma(y)}\right] \partial_\mu \partial_\nu r(x) \label{gaugederiv}  \\ 
h^{new}_{\mu 5}(x,y) &=& 0  \\
h^{new}_{55}(x,y) &=& \alpha_0\ r(x)  
\eea
where $\alpha_0$ is given in eq. (\ref{constants}) and $C_0$ is an unphysical integration constant.

The metric ansatz that we get in this new gauge is the same as the one proposed in \cite{boos} (they write $\ \phi(x)=\alpha_0 r(x)\ $ and choose $\ C_0 = 0\ $), which was introduced to solve possible problems of the `bent-brane' procedure used for example in \cite{rubakov,garriga}. As it now becomes clear, both aproaches are correct and equivalent to linear order, being related by the gauge transformation defined above. 
 We stress that our ansatz was derived without any `bent brane' procedure, but is equal to the linear expansion of the metric solution of \cite{rubakov}.

We also need to mention that in \cite{das}, a complete investigation on the general features of the RS scenario, another result for the metric perturbations is given and seems apparently incompatible with ours. But, there are only two differences. The first one resides in that they choose to start in the same gauge used in \cite{boos} (i.e., the gauge defined in this section, although not necessarilly with $C_0=0$). The second is their method to solve the coupled equations, which involved expanding first all the fields in complete sets of functions of the y-coordinate and then decoupling the equations. Because they start in the same gauge (up to the constant $C_0$), their analysis should simply be equivalent to the one in \cite{boos}, and, therefore, to ours. 

A simpler gauge transformation (starting in our gauge) in which $\xi_5 =0$ and \\
$\xi_\mu = -{1\over 2} C_0\ e^{-2\sigma(y)} \partial_\mu r(x)$ would give
\bea
h^{new}_{\mu\nu}(x,y) &=& h_{\mu\nu}(x,y)\ -\  \eta_{\mu\nu}\   r(x)  + C_0 e^{-2\sigma(y)} \partial_\mu \partial_\nu r(x)  \\ 
h^{new}_{\mu 5}(x,y) &=& 0  \\
h^{new}_{55}(x,y) &=& 2 e^{2 \sigma}  r(x)  
\eea
which was an ansatz discussed in \cite{pilo}, also gauge equivalent to our ansatz.

\section{Conclusions}

We have worked out the linear metric perturbations in the context of the RS1 setup. To simplify the task, we fixed a gauge, in which the two branes remained at fixed y-position, and reduced the number of metric perturbations. Einstein equations turned out to couple the remaining physical fields, but we carried out a very simple decoupling procedure and we ended up with the equations of motion of a massless scalar, the radion, and a 4-dimensional 2 tensor field, verifying transverse-traceless conditions.

The metric ansatz resulting from this analysis turned out to be equal to the linearly expanded ansatz given in \cite{rubakov} where the `bent brane' method was used. We also showed that our ansatz is completely equivalent to the more complicated ansatz calculated in \cite{boos}, the difference being only on the gauge choice and, therefore, unphysical. We also argued that the ansatz of \cite{das} should only differ with ours by the choice of gauge, which is (up to a constant $C_0$) the same as \cite{boos}.  

We believe to have solved an apparent conflict between `different' metric solutions that turned out to be all equivalent and correct, being simply written down in different but compatible 5-dimensional coordinate frames.

We did not discuss any stabilization issues and their influence on the metric (backreaction), although they are evidently of great interest and will be the subject of future studies.

\label{concl}
\vspace{.2cm}

\centerline{\bf \Large Acknowledgements}

\vspace*{1cm}

I would like to thank James Wells, John Gunion, Steve Carlip, Bohdan Grzadkowski and Sayan Basu for their patience and helpful conversations. I would also like to thank the hospitality of Ch\'erif Hamzaoui and the Physics Department of UQ\`AM, in Montr\'eal. Special thanks too, to the TASI summer school where this project took its final momentum.

\vspace*{1cm}


\end{document}